\documentstyle[prl,aps,floats,epsf]{revtex}
\begin{document}
\twocolumn[\hsize\textwidth\columnwidth\hsize\csname@twocolumnfalse%
\endcsname

\draft

\title{\Large\bf Quantum Heisenberg Antiferromagnets  \\
versus Nonlinear $\sigma$ Model Without the Large $S$ Limit}

\author{\bf Wei-Min Zhang}

\address{Department of Physics, National Cheng Kung
University, Tainan, 701, Taiwan}

\date{\today}
\maketitle

\begin{abstract}
{In this letter, I develop a new topologically
invariant coherent state path integral for spin systems,
and apply it to the quantum Heisenberg model on a square
lattice. As a result, the quantum nonlinear $\sigma$ model
for arbitrary values of
spin can be directly obtained. The effective coupling
constant and spin wave velocity are modified by $g_s=
{2\over S}\sqrt{d+{T_\Lambda\over 2SJ}}$ and $c_s=2JSa
\sqrt{d+{T_\Lambda\over 2SJ}}$, where $T_\Lambda$
is a natural temperature scale for the reliability of
the theory. The formulation can also be extended to other
generalized coherent state path integrals.}
\end{abstract}
%
]

Physics of quantum Heisenberg antiferromagnets (QHA) in
low dimensional strongly correlated systems continuously
attract attentions. This is largely due to the intense
interests in understanding high $T_c$ copper-oxide
superconductivity which arises as a consequence of hole
(or electron) dopings from the parent antiferromagnetic
(AF) Mott insulators \cite{mak98}. The AF Mott insulators
concerned in high $T_c$ basically correspond to a
two-dimensional spin-${1\over 2}$ lattice QHA with
nearest-neighbor exchange interaction, which is, quantal
mechanically, still a difficult problem to solve.
Meanwhile, there are also increasing interests in
quasi-one-dimensional Heisenberg spin chains because
of the stripe phase evidenced recently from the observed
incommensurate magnetic excitations in cuprate
superconductors\cite{stripe}.
In the {\it large spin $S$ limit}, Haldane has shown
that the lattice QHA can be described
by an quantum nonlinear $\sigma$ model (QNL$\sigma$M)
when the correlation length is vary large, i.e. {\it
at low temperatures}\cite{haldane}. Since then, the
QNL$\sigma$M has become a good candidate to describe
{\it phenomenologically} various experimental data of the
two-dimensional QHA \cite{Chak89,HN91,Chub94}.

However, two crucial questions arise for some time:
why the predictions of the long-wavelength, low
energy physics of the QNL$\sigma$M obtained in large $S$
limit can truely be identical to experimental data of
the lattice QHA in vary small values of $S$? and which is
the actual temperature upper limit where the QNL$\sigma$M
is reliable? In fact, it has been found that the correlation
length $\xi(T)$ derived from the QNL$\sigma$M \cite{HN91}
is not always agreement with the experimental data of
the QHA\cite{Cuccoli}. Also, the temperature dependence
of $\xi(T)$ does not occur the so-called quantum critical
regime predicted by QNL$\sigma$M \cite{Car20}.
In this letter, by improving the formalism of spin
coherent state path integral, I derive the QNL$\sigma$M
form the lattice QHA for arbitrary values of spin.
Meanwhile, a temperature scale $T_\Lambda$ for the
reliability of the theory is naturally realized.

The general spin coherent state path integral is known
for a long time\cite{wzhang90}. Explicitly, the partition
function can be expressed (in units $\hbar=k_B=1$) by 
\begin{equation} \label{cspi}
   Z= Tr e^{-\beta H}=\int {\cal D}[\bbox{\Omega}]e^{
        \int_0^\beta d\tau \Big\{i S {\bf A}
        \cdot \dot{\bbox{\Omega}}-\langle \bbox{\Omega} |
        H |\bbox{\Omega} \rangle \Big\}}
\end{equation}
where the term $iS \int_0^\beta d\tau {\bf A} \cdot
\dot{\bbox{\Omega}}=\int_0^\beta d\tau\langle \bbox{\Omega}|
{d\over d\tau}|\bbox{\Omega} \rangle \equiv iS\omega(\bbox{\Omega})$
is a topological Berry phase at the site $i$\cite{Wilczek},
and ${\bf A}$ is a U(1) monopole potential, the state
$|\bbox{\Omega}\rangle$ is a spin coherent state
which can be constructed by rotating the ``north pole''
state $|S,S\rangle$ to along the $\bbox{\Omega}$
direction, $|\bbox{\Omega}\rangle = R(\bbox{\Omega})
|S,S\rangle$. In other words, $\bbox{\Omega}$ is a unit
vector along which the spin operator with quantum number
$S$ is maximally aligned in $|\bbox{\Omega}\rangle$.
For the lattice Heisenberg model, $H=J\sum_{<ij>}{\bf S}_i
\cdot {\bf S}_j$ $(J>0)$ which has the global SO(3) spin
rotational symmetry, one can obtain
\begin{equation} \label{hmpi}
    Z_H =\int {\cal D}[\bbox{\Omega}_i]e^{
        \int_0^\beta d\tau \Big\{iS\sum_i {\bf A}_i \cdot
        \dot{\bbox{\Omega}}_i- JS^2 \sum_{<ij>}
        \bbox{\Omega}_i \cdot \bbox{\Omega}_j\Big\} } .
\end{equation}
The whole exponent in Eq.~(\ref{hmpi}) represents an effective
action of the Heisenberg model in terms of the spin coherent
state path integral. By minimizing $H(\bbox{\Omega})=
JS^2\sum_{<ij>}\bbox{\Omega}_i\cdot \bbox{\Omega}_j$, one
can find the classical ground state (N\'{e}el state) which
spontaneously breaks the SO(3) symmetry. Then expanding the
action around the ground state, one can easily derive the
spin-wave theory for the QHA that
describes the long wavelength spin modes\cite{Auerbach}.

However, the spin wave theory is only applicable for
long-range ordered phase where the SO(3) symmetry is
spontaneously broken. Based on Mermin-Wagner's
theorem\cite{mwt}, no symmetry can be spontaneously
broken for $T>0$ in one and two dimensional Heisenberg
model. In other words, the spin wave theory is invalid
in low dimensions. To derive an effective long
wavelength action that keeps the full spin rotational
symmetry, Haldane considered the large $S$ limit.
In the large $S$ limit, the path integrals of
Eq.~(\ref{hmpi}) are dominated by the semiclassical
equation: $iS\bbox{\Omega}\times\dot{\bbox{\Omega}}
= {\partial H[\bbox{\Omega}]\over \partial
\bbox{\Omega}}$. By separating the semiclassical solution
$\bbox{\Omega}_i$ into a slowly varying N\'{e}el order
unit vector $(-1)^i {\bf n}(x_i)$ plus a slowly varying
magnetization density field perpendicular to ${\bf n}(x_i)$
(Haldane mapping), then taking the continuous limit and
integrating out the magnetic density field, Haldane shows
that Eq.~(\ref{hmpi}) can be reduced to the QNL$\sigma$M,
\begin{equation} \label{nlsm}
   Z_H \propto \int {\cal D}[{\bf n}]~ e^{i2\pi S\Theta[{\bf n}]}
        ~e^{-{\Lambda^{d-1}\over 2g_0}\int d^{d+1}x~
            \partial_\mu {\bf n} \partial^\mu {\bf n}},
\end{equation}
which is defined in the $d$+1-dimensional space $(x^1$, $\cdots,
x^{d+1})$=$(x^1, \cdots, x^d, c_o\tau)$, where $d$ represents
the $d$-dimensional lattice space of the Heisenberg model,
$\Lambda=a^{-1}$ is an intermediate momentum cutoff as the
inverse of the lattice spacing $a$, $g_o=c_o a^{1-d}/\rho^o_s
=2\sqrt{d}/S$ the dimensionless coupling constant,
$c_o=2\sqrt{d}Jsa$ the spin wave velocity, and
$\rho^o_s=JS^2a^{2-d}$ the spin stiffness. The imaginary time
(temperature) variable $\tau$ ranges from $0$ to $ \beta
=1 /T$. The exponent $\Theta[{\bf n}]$ in Eq.~(\ref{nlsm})
is a topological factor associated with the Berry phase,
which distinguishes between integer and half-integer spins.

In this letter, by improving the formulation (\ref{cspi})
of the spin coherent state path integral, I can indeed
directly obtain the QNL$\sigma$M by integrating out the
quantum fluctuation above the energy scale $k_BT_\Lambda$.
In this derivation, I neither assume the large $S$
limit nor use Haldane's mapping so that the result is
applicable to arbitrary values of spin. Meanwhile, it
also solves a long-standing problem in the construction
of generalized coherent state path integrals\cite{Schu}.

Simply speaking, the long-standing problem in the
construction of coherent state path integrals
arises from the fact that Eq.~(\ref{cspi})
is not well defined, so does Eq.~(\ref{hmpi}).
The main suspect comes
from the assumption that $|\bbox{\Omega}(\tau+\delta\tau)
\rangle-|\bbox{\Omega}(\tau)\rangle$ is order of
$O(\delta\tau)$ used in the derivation of Eq.~(\ref{cspi}).
Although this assumption has been widely accepted in all
the applications of generalized coherent state path
integrals, it has never been justified\cite{Schu}.
As Klauder first pointed out\cite{Klauder}, the semiclassical
(or stationary phase) approximation of Eq.~(\ref{cspi})
[or (\ref{hmpi}) that Haldane used to derive Eq.~(\ref{nlsm})]
is indeed incompatible with the required initial and final
boundary conditions embedded in the coherent state
path integral.

Explicitly, coherent state path
integrals are defined on a geometric space $G/H$ (here is
$SU(2)/U(1) \sim S^2$) which has a phase space structure
that the curvature $d\omega(\bbox{\Omega})$ obtained from
the Berry phase serves as a symplectic structure for the
corresponding semiclassical motion\cite{wzhang90}. Then,
for each pair $(p,q)$ which obeys two first-order time
differential equations, there exist two pair (i.e. four)
boundary conditions $(p_0,q_0)$ and $(p_f,q_f)$ from the
initial and final fixed points, which is
incompatible. In the literature, there are two possible
ways to avoid this inconsistency. One approach was pointed
out by Faddeev\cite{Faddeev} that one must specify the initial
and final boundary conditions only by the independent
complex variable $z_i$ and $z^*_f$ respectively due to
the complex structure of quantum state\cite{wzhang90}.
Another approach is {\it proposed} by Klauder that one
may add an additional square term of time-derivative,
$(\dot{\bbox{\Omega}})^2$ into
the action of Eq.~(\ref{cspi}) to match the additional
initial and final boundary conditions in phase
space\cite{Klauder}.

Physically, the above inconsistency
arises essentially from the ignoring quantum fluctuations
in the coherent state action. In the path integral
formulation, there always exist simultaneously
fast and slow varying paths
that are associated with short and long range quantum
fluctuations, respectively. The effective action
for slow varying motions can be properly obtained by
integrating over short range quantum fluctuations.
However, in the derivation of Eq.~(\ref{cspi}), only
these slowly varying motions are concerned and short
range quantum fluctuations are simply dropped from
the off-diagonal matrix elements.
This causes the incompatibility between the semiclassical
approximation and the boundary conditions.

To see more explicitly, we begin with the discrete form of the
partition function derived exactly from the coherent state
representation\cite{wzhang90}
\begin{eqnarray} \label{dispi}
    Z=\lim_{N\rightarrow \infty} &&\prod_{k=1}^{N}d\mu(
        \bbox{\Omega}^k) \exp \Big\{\sum_{k=1}^N \ln
        \langle \bbox{\Omega}^k|\bbox{\Omega}^{k-1}
        \rangle \nonumber \\
    &&~~~~~~~~~~~~~~~~~~~~~~ -\epsilon {\langle
        \bbox{\Omega}^k | H | \bbox{\Omega}^{k-1}\rangle
        \over \langle \bbox{\Omega}^k|\bbox{\Omega}^{k-1}
        \rangle} \Big\},
\end{eqnarray}
where $|\bbox{\Omega}^N\rangle=| \bbox{\Omega}^0\rangle$
because of the periodicity of the trace),
$\epsilon=\beta/N$ is infinitesimal as $N\rightarrow
\infty$. The assumption that $|\bbox{\Omega}^k\rangle-
|\bbox{\Omega}^{k-1}\rangle$ is order of $0(\epsilon)$
only extracts slowly varying motions, where
rapidly varying motions are neglected. As a result,
the time continuous limit of (\ref{dispi})
is just the conventional spin
coherent state path integral given by Eq.~(\ref{cspi}).

To include the contribution
of rapidly varying paths, one should expand the near-by
coherent state overlap to the second order terms that
either are exclusively slowly varying or include at least
one rapidly vary term. The topologically invariant terms
of such contributions can be uniquely expressed by
\begin{eqnarray} \label{geom}
     && \lim_{N\rightarrow \infty} \sum_{k=1}^N \ln \langle
        \bbox{\Omega}_k|\bbox{\Omega}_{k-1} \rangle =
        S\int_0^\beta d\tau \Big\{i {\bf A} \cdot
        \dot{\bbox{\Omega}} \nonumber \\
     && ~~~~~~~~ + i\bbox{\Omega}\cdot
        (\dot{\bbox{\Omega}}\times \delta \bbox{\Omega})
        - {1\over \tau_\Lambda}\delta \bbox{\Omega}\cdot
        \delta \bbox{\Omega} + \cdots \Big\} ,
\end{eqnarray}
where the time derivative $\dot{\bbox{\Omega}}$ and the
variation $\delta\bbox{\Omega}$ represent slowly varying
motions and short range fluctuations (i.e., rapidly
varying motions), respectively. The parameter $\tau_\Lambda$
is an intrinsic shortest timescale to distinguish between
slowly varying and rapidly varying motions, which I will
discuss in details later. The second and third terms
in Eq.~(\ref{geom}) are usually neglected in the
conventional treatment of path integrals. For the
Hamiltonian term, since it is already proportional
to $\epsilon$, I only keep the off-diagonal expansion
up to the quadratic order of $\delta \bbox{\Omega}$:
\begin{eqnarray}
\lim_{N \rightarrow \infty} && \sum_{k=1}^N \epsilon
    {\langle\bbox{\Omega}^k | H | \bbox{\Omega}^{k-1}
    \rangle \over \langle \bbox{\Omega}^k|
    \bbox{\Omega}^{k-1}\rangle} = \int d\tau
    \Big\{H[\bbox{\Omega}] \nonumber \\
& & + {\partial H[\bbox{\Omega}]\over \partial
    \bbox{\Omega}} \cdot \delta \bbox{\Omega} +
    {\partial^2 H[\bbox{\Omega}]\over
    \partial \bbox{\Omega}_\alpha\partial
    \bbox{\Omega}_{\alpha'}} \delta \bbox{\Omega}_\alpha
    \delta\bbox{\Omega}_{\alpha'} + \cdots \Big\} .
    \label{dyna}
\end{eqnarray}
where $\alpha, \alpha'$ are indices of spin components.
Substituting Eqs.~(\ref{geom},\ref{dyna}) into
Eq.~(\ref{dispi}), one has
\begin{eqnarray}
Z=\int && {\cal D}[\bbox{\Omega}] {\cal D}[\delta
    \bbox{\Omega}]~ \exp \int_0^\beta d\tau
    \Big\{iS{\bf A} \cdot \dot{\bbox{\Omega}}
    -H[\bbox{\Omega}] \nonumber \\
&& ~~~~~~~~ + \Big[ iS\bbox{\Omega}\times
    \dot{\bbox{\Omega}}-{\partial H[\bbox{\Omega}]
    \over \partial \bbox{\Omega}}
    \Big] \cdot \delta \bbox{\Omega} \nonumber \\
&& ~~ - \Big[{S\over \tau_\Lambda}
    \delta_{\alpha{\alpha'}} + {\partial^2
    H[\bbox{\Omega}]\over \partial
    \bbox{\Omega}_\alpha \partial
    \bbox{\Omega}_{\alpha'}}\Big] \delta
    \bbox{\Omega}_\alpha \delta
    \bbox{\Omega}_{\alpha'} + \cdots \Big\} , \label{cspie}
\end{eqnarray}
which describes both the slowly varying
motion $\dot{\bbox{\Omega}}$
and the short range fluctuations $\delta{\bbox\Omega}$.

If one artificially drops the dynamical dependence of
short range fluctuations [namely, the higher order terms
originated from the off-diagonal matrix element of the
Hamiltonian in Eq.~(\ref{dyna})], and integrate out
the remaining $\delta\bbox{\Omega}$ terms in
Eq.~(\ref{cspie}), it turns out that:
\begin{equation}
    Z=\int {\cal D}[\bbox{\Omega}]
        ~ \exp \int_0^\beta d\tau \Big\{iS{\bf A}
        \cdot \dot{\bbox{\Omega}}-H[\bbox{\Omega}]
        - {1\over 4}S\tau_\Lambda \dot{\bbox{\Omega}}^2\Big\}.
        \label{klaud}
\end{equation}
This is just Klauder's nonconventional coherent state path
integral\cite{Klauder}.
As one can see this term is originated of integrating
over short range quantum fluctuations. Note that
the shortest timescale $\tau_\Lambda={1\over T_\Lambda}$.
It shows that the effective contribution of short
range fluctuations is obtained by integrating over
quantum dynamics above the temperature (energy)
$T_\Lambda$. When $T_\Lambda \rightarrow \infty$,
this term vanishes. Namely, no short-range
fluctuations are renormalized, as one expected\cite{Wilson}.
Hence, the coefficient of this topologically invariant
metric term, i.e. $T_\Lambda$, serves as a factorization
temperature scale between long and short range
quantum fluctuations. This temperature scale (or
time scale in real-time dynamics) must be determined
from the dynamics of the original Hamiltonian.
In other words, naively ignoring the dynamical dependence
of short range fluctuations in (\ref{cspie}) is
physically not meaningful. Therefore, only in the absence
of a Hamiltonian, Klauder's formulation is correct.

On the other hand, it is interesting to see that
if one requires vanish of the second term,
$iS\bbox{\Omega}\times\dot{\bbox{\Omega}} -
{\partial H[\bbox{\Omega}]\over \partial\bbox{\Omega}}=0$,
Eq.~(\ref{cspie})
is just a variation expansion of (\ref{cspi}) with
respect to the semiclassical dynamics Haldane used\cite{haldane}.
However, Eq.~(\ref{cspie}) here is derived by carefully
treating the off-diagonal matrix elements of near-by coherent
states in Eq.~(\ref{dispi}) in terms of the short range
fluctuation $\delta\bbox{\Omega}$ and the slowly varying motions
$\dot{\bbox{\Omega}}$. There is no semiclassical expansion
beginning with. Hence, it has no necessary requirement to
take the semiclassical limit by letting the second term vanish.
Instead of, one shall integrate out the short range fluctuations
$\delta{\bbox{\Omega}}$ to obtain a low energy effective
action that can describe the long wavelength spin modes.
Since no semiclassical approximation is made, the resulting
low energy effective action is valid for arbitrary values
of spin.

Now, we can apply the new formulation of spin coherent state
path integral (\ref{cspie}) to the square lattice QHA.
To specify the AF ordering, let the slowly varying $\bbox{\Omega}_i
=(-1)^i{\bf n}(x_i)$, here the Ne\'{e}l order ${\bf n}(x_i)$
is a unit vector $|{\bf n}(x_i)|=1$. Then taking the
space continuous limit $\sum_i \rightarrow {1\over a^d}\int
d^d x$ where $a$ is the lattice spacing:
\begin{eqnarray}
    &&H[\bbox{\Omega}] = JS^2 \sum_{<ij>}
    \bbox{\Omega}_i \cdot \bbox{\Omega}_j \nonumber \\
    &&~~\rightarrow -dJS^2N + {JS^2\over 2a^{d-2}} \int d^d x
    \sum_{k=1}^d[\partial_k {\bf n}(x)\cdot \partial_k {\bf n}(x)],
    \label{h1} \\
    && {\partial H[\bbox{\Omega}]\over \partial \bbox{\Omega}}
    \cdot \delta \bbox{\Omega} \rightarrow 0 , \label{csc} \\
    &&{\partial^2 H[\bbox{\Omega}]\over \partial \bbox{\Omega}_\alpha
    \partial \bbox{\Omega}_\beta} \delta \bbox{\Omega}_\alpha \delta
    \bbox{\Omega}_\beta \rightarrow {2dJS^2 \over a^d} \int d^d x
    \delta \bbox{\Omega}(x)\cdot\delta \bbox{\Omega}(x). \label{h3}
\end{eqnarray}
Substituting (\ref{h1}-\ref{h3}) into (\ref{cspie}) and
integrating out the short range fluctuation
$\delta{\bbox{\Omega}}$, I obtain,
\begin{eqnarray}
    &&Z_H \propto \int {\cal D}[{\bf n}]~ e^{i2\pi S\Theta[{\bf n}]}
        ~\exp\Big\{- {a^{1-d}\over 2g_s} \int^{c_s\over
        T}_{c_s\over T_{\Lambda}}d(c_s\tau) \nonumber \\
    &&~~~~~~~~~~~~~~~~~~~\times  \int_a d^d x~\Big[{1\over c_s^2}
        \Big|{\partial{\bf n}\over \partial \tau}\Big|^2
        + |\nabla_x {\bf n}|^2 \Big] \Big\}. \label{qnlsm}
\end{eqnarray}
This is the QNL$\sigma$M for low energy QHA with arbitrary
values of spin. Note that two basic parameters, the dimensionless
coupling constant $g_s$ and the spin wave velocity $c_s$,
in (\ref{qnlsm}) crucially depend on $S$ and $T_\Lambda$:
\begin{equation}
    g_s={2\over S} \sqrt{d+{T_\Lambda\over 2SJ}}~,~~
    c_s=2JSa\sqrt{d+{T_\Lambda\over 2SJ}}~.  \label{bsc}
\end{equation}
While, the topological phase factor $2\pi S \Theta[{\bf n}]=
2\pi S\sum_i(-1)^i{\bf A}(x_i)\cdot\dot{\bf n}(x_i)$
keeps the same as in Haldane's derivation.

The difference between (\ref{nlsm}) and (\ref{qnlsm})
mainly comes from the contribution of the pure quantum fluctuation,
i.e., the $1/\tau_\Lambda$ term in (\ref{cspie}) which
is ignored in Haldane's mapping for the large $S$ limit
\cite{haldane} but it plays an important as argued
by Klauder\cite{Klauder}. Also, this difference is indeed
associated with the temperature scale for the reliability
of the QNL$\sigma$M. Usually one thinks that there should
be no intrinsic cutoff for the imaginary time variable
$\tau$ because quantum fluctuations exist on all time scale
in path integrals\cite{Chak89}. But a low energy effective
theory constructed from path integral is defined by
integrating over high energy dynamics,
i.e., high energy quantum fluctuations
above some energy scale $T_\Lambda$. Without such an
intrinsic cutoff, namely, let $1/\tau_\Lambda \rightarrow 0$
or $T_\Lambda \rightarrow \infty$, Eq.~(\ref{qnlsm}) is
reduced to
\begin{equation}
    Z_H \propto \int {\cal D}[{\bf n}]~ e^{i2\pi S\Theta[{\bf n}]}
        ~\exp\Big\{- {\rho_s\over 2T} \int_a d^d x~
        |\nabla_x {\bf n}|^2 \Big] \Big\}. \label{qnlsm1}
\end{equation}
where $\rho_s=JS^2a^{2-d}$ is the spin stiffness. Except for
the topological phase, this is just the classical
$d$-dimensional NL$\sigma$M rather than Haldane's
$d+1$-dimensional NL$\sigma$M\cite{haldane}. That is,
without using this intrinsic short-time cutoff or naively
ignore this cutoff, it should be hard to correctly extract
the fundamental quantum effect of QHA.

On the other hand,
the lattice spacing $a$ indicates the existence of an intrinsic
momentum cutoff $\Lambda$ in the $d$-dimensional momentum
space: $\Lambda=2\sqrt{\pi}[\Gamma(d/2+1)]^{1/d}/a\equiv L/a$.
Correspondingly, there must exist an intrinsic energy cutoff
$T_\Lambda=c_s\Lambda/2\pi$\cite{Auerbach}. Combining with
(\ref{bsc}), it turns out that
\begin{equation} \label{scale}
    {T_\Lambda \over J}={SL^2\over 4\pi^2}
        \Big(1+\sqrt{1+{16\pi^2 d\over L^2}}~\Big) .
\end{equation}
For $d=2$ and $S=1/2$, we have $L=2\sqrt{\pi}$ and thus $T_\Lambda/J
\simeq 0.97$. This provides quantitatively a temperature upper
limit for the reliability of the QNL$\sigma$M:
\begin{equation}
   0 \leq T/J < T_\Lambda/J \simeq 1.0 ~.
\end{equation}
Meanwhile, the spin wave velocity $c_s$ can also be explicitly
determined from (\ref{bsc}) and (\ref{scale}). For La$_2$CuO$_4$
which is a typical $d=2$ spin-1/2 QHA with
$a=3.79$\AA~and $J \simeq 1500 K$, I obtain (inset back
the unit $\hbar$)
\begin{equation}
    \hbar c_s=2JSa\sqrt{d+{T_\Lambda\over 2SJ}}
        \simeq 0.85~ {\rm eV ~\AA}~.
\end{equation}
This is in excellent agreement with the experimental data
$\hbar c_s = 0.85\pm 0.03$ eV \AA \cite{Aeppli}.

Now one can see that the basic parameters in QNL$\sigma$M,
i.e. the coupling constant $g_s$ and the spin wave velocity
$c_s$, are unambiguously defined in terms of $J, S, a$ and
$d$ in the spin system. Let $y_i=x_i/a (i=1, \cdots, d)$,
$y_{i+1}=c_s\tau/a$, Eq.~(\ref{qnlsm}) becomes
\begin{eqnarray}
    &&Z_H \propto \int {\cal D}[{\bf n}]~ e^{i2\pi S\Theta[{\bf n}]}
        ~\exp\Big\{- {1\over 2g_s} \int^{{T\over T_\Lambda}
        2\pi/L}_{2\pi/L}dy_{i+1} \nonumber \\
    &&~~~~~~~~~~~~~~~~~~~~~~\times  \int_1 d^d y~ \sum_{\mu=1}^{d+1}
        \partial_{\mu} {\bf n} \partial^{\mu} {\bf n} \Big\}.
\end{eqnarray}
which is a dimensionless QNL$\sigma$M, where $L$ is only a
geometrical constant
[see above Eq.~(\ref{scale})]. The temperature dependence of the
coupling constant $g_s$ is given by the running coupling
constant $g_s(T)$ which can be determined by the standard
renormalization group approach\cite{Chak89,HN91}. Further
applications of our QNL$\sigma$M to thermodynamic
properties of the QHA for $T<T_\Lambda$,
i.e. the calculations of internal energy,
specific heat, correlation functions, staggered susceptibility,
and correlation length, are straightforward for different values
of spin and lattice dimension. The effect of topological
phase that distinguishes between integer and half-integer
spins can also be studied. The results will be presented in
separate publications. Moreover, The construction of a low
energy effective field theory from the extended coherent
state path integrals developed in this letter is a general approach, in
which the use of the shortest timescale plays an important
role in order to obtain a self-consistent effective field
theory. This approach can also be applied to other generalized
coherent state path integrals\cite{wzhang90} for the study of
strongly correlated or strongly interacting systems.

This work is supported by NSC 89-2112-M-006-029.



\begin{references}
    \bibitem{mak98} For a review, see M. A. Kastner et al.,
        {\it Rev. Mod. Phys.} {\bf 70}, 897 (1998).
    \bibitem{stripe} H. A. Mook et al, {\it Nature}, {\bf 395},
        580 (1998);
    \bibitem{haldane} F. D. M. Haldane, {\it Phys. Rev. Lett.}
        {\bf 50}, 1153 (1983); {\it Phys. Lett.} {\bf 93A},
        464 (1983).
    \bibitem{Chak89} S. Chakravarty, B. I. Halperin and D. R.
        Nelson, {\it Phys. Rev. Lett.} {\bf 60}, 1057 (1988);
        {\it Phys. Rev.} {\bf B 39}, 2344 (1989).
    \bibitem{HN91} P. Hasenfratz and F. Niedermayer, {\it Phys.
        Lett.} {\bf B 268}, 231 (1991).
    \bibitem{Chub94} A. V. Chubukov, S. Sachdev and J. Ye, {\it Phys.
        Rev.} {\bf B 49}, 11919 (1994).
    \bibitem{Cuccoli} A. Cuccoli et al. {\it Phys. Rev. Lett.}
        {\bf 77}, 3439 (1996).
    \bibitem{Car20} P. Carretta et al., {\it Phys. Rev. Lett.}
        {\bf 84}, 366 (2000).
    \bibitem{wzhang90} W. M. Zhang, D. H. Feng
        and R. Gilmore, {\it Rev. Mod. Phys.} {\bf 62}, 867 (1990).
    \bibitem{Wilczek} A. Shapere and F. Wilczek, {\it Geometric Phase
        in Physics}, (World Scientific, Singapore, 1989).
    \bibitem{Auerbach} A. Auerbach, ``Interacting Electrons and
        Quantum Magnetism'', (Springer-Verlag, 1994).
    \bibitem{mwt} N. D. Mermin and H. Wagner, {\it Phys. Rev. Lett.}
        {\bf 17}, 1133 (1966).
    \bibitem{read}R. Shankar and N. Read, {\it Nucl. Phys.}
        {\bf B 336}, 457 (1990).
    \bibitem{Schu} L. S. Schulman, {\it Techniques and Applications
        of Path Integration}, (Wiley, New York, 1981);
    \bibitem{Klauder} J. R. Klauder, {\it Phys. Rev. } {\bf D 19}, 2349
        (1979).
    \bibitem{Faddeev} L. D. Faddeev, in {\it Methods in Field Theory},
        Les Houches 1975, Ed. by R. Balian and J. Zinn-Justin
        (North-Holland, Amsterdam, 1976) p.1
    \bibitem{Wilson} K. G. Wilson and J. Kogut, {\it Phys. Rep.}
        {\bf 12C}, 75 (1974).
    \bibitem{Aeppli} G. Aeppli et al. {\it Phys. Rev. Lett.} {\bf 62},
        2052 (1989)
\end{references}
\end{document}